\documentclass[aip,jap,reprint,graphicx]{revtex4-1} 

\usepackage{graphicx}
\usepackage{dcolumn}
\usepackage{bm}
\usepackage{bbm}
\usepackage{hyperref}

\begin{document}

\title{NV-center imaging of bubble domains in a 6-\AA~film of cobalt with perpendicular magnetization}

\author{J.-P.~Tetienne}
\email{jtetienn@ens-cachan.fr}
\author{T.~Hingant}
\affiliation{Laboratoire Aim\'e Cotton, CNRS, Universit\'e Paris-Sud and ENS Cachan, 91405 Orsay, France}
\affiliation{Laboratoire de Photonique Quantique et Mol\'eculaire, ENS Cachan and CNRS UMR 8537, 94235 Cachan, France}
\author{L.~Rondin}
\affiliation{Laboratoire de Photonique Quantique et Mol\'eculaire, ENS Cachan and CNRS UMR 8537, 94235 Cachan, France}
\author{S.~Rohart}
\author{A.~Thiaville}
\affiliation{Laboratoire de Physique des Solides, Universit\'e Paris-Sud and CNRS UMR 8502, 91405 Orsay, France}
\author{E.~Ju\'e}
\affiliation{SPINTEC, UMR 8191, CEA/CNRS/UJF/GINP, INAC, 38054 Grenoble, France}
\author{G.~Gaudin}
\affiliation{SPINTEC, UMR 8191, CEA/CNRS/UJF/GINP, INAC, 38054 Grenoble, France}
\author{J.-F.~Roch}
\affiliation{Laboratoire Aim\'e Cotton, CNRS, Universit\'e Paris-Sud and ENS Cachan, 91405 Orsay, France}
\author{V.~Jacques}
\affiliation{Laboratoire Aim\'e Cotton, CNRS, Universit\'e Paris-Sud and ENS Cachan, 91405 Orsay, France}
\affiliation{Laboratoire de Photonique Quantique et Mol\'eculaire, ENS Cachan and CNRS UMR 8537, 94235 Cachan, France}

\begin{abstract}

We employ a scanning NV-center microscope to perform stray field imaging of bubble magnetic domains in a perpendicularly magnetized Pt/Co/AlO$_x$ trilayer with 6 \AA~of Co. The stray field created by the domain walls is quantitatively mapped with few-nanometer spatial resolution, with a probe-sample distance of about 100 nm. As an example of application, we show that it should be possible to determine the Bloch or N{\'e}el nature of the domain walls, which is of crucial importance to the understanding of current-controlled domain wall motion.

\end{abstract}

\maketitle

Ultrathin ferromagnets have attracted considerable interest over the last years due to their potential use in low power spintronic devices. A typical example of such magnets is the Pt/Co/AlO$_x$ trilayer, where the ferromagnetic Co layer can be as thin as a few atomic planes and generally exhibits perpendicular anisotropy. Evidence of fast current-induced domain wall (DW) motion in this system~\cite{Miron2010} has initiated numerous studies on ultrathin magnetic layers sandwiched in an asymetric stack. Such systems have been proven to exhibit spin orbit torques~\cite{Miron2010,Miron2011,LiuScience2012,Garello2013}, which are believed to play a key role in the DW dynamics \cite{Thiaville2012,Ryu2013, Emori2013}. This interpretation relies on the modification of the DW structure by the Dzyaloshinskii-Moriya interaction, from the well-known Bloch type expected in such thin layer with perpendicular anisotropy to the N\'eel type. However, a clear evidence of the DW structure, whether N\'eel or Bloch type, is still missing. Indeed, there is a lack of experimental techniques enabling direct imaging of such DWs with nanoscale spatial resolution. Techniques based on beams of X-rays~\cite{Heyne2008,Im2009,Vogel2012} or electrons~\cite{Klaui2005,Biziere2013} interacting with the sample suffer from a lack of signal owing to the small interaction volume, while magnetic force microscopy~\cite{Yamaguchi2004} is usually not suitable because of the high sensitivity of DWs in ultrathin films to magnetic perturbations. Finally only two techniques used for model samples, spin-polarized scanning tunelling microscopy~\cite{Meckler2009} and spin-polarized low energy electron microscopy~\cite{Chen2013}, have proven their ability to image the DW structure. However, samples commonly used in spintronics devices require more flexible imaging techniques.\\
\indent In this letter, we report direct imaging of DWs in Pt/Co/AlO$_x$ with 6 \AA~of Co by using a single nitrogen-vacancy (NV) defect in diamond as magnetic sensor. Scanning NV-center microscopy is a recently introduced magnetometry technique \cite{Taylor2008,Balasubramanian2008} that allows quantitative mapping of the stray magnetic field~\cite{Rondin2012,Maletinsky2012}, with high sensitivity \cite{Maze2008}, no significant magnetic back-action on the sample, and a spatial resolution ultimately limited by the atomic size of the probe. Importantly, it can operate under ambient conditions and on nearly any kind of samples. Recent striking results include stray field imaging of a single electron spin~\cite{Grinolds2013}, of the vortex core in a microdot of permalloy~\cite{Rondin2013,Tetienne2013}, and of living magnetotactic bacteria~\cite{LeSage2013}. In this work, we apply NV-center microscopy to image bubble magnetic domains in a continuous Pt/Co/AlO$_x$ film with perpendicular magnetization.\\
\begin{figure}[t]
\begin{center}
\includegraphics[width=0.49\textwidth]{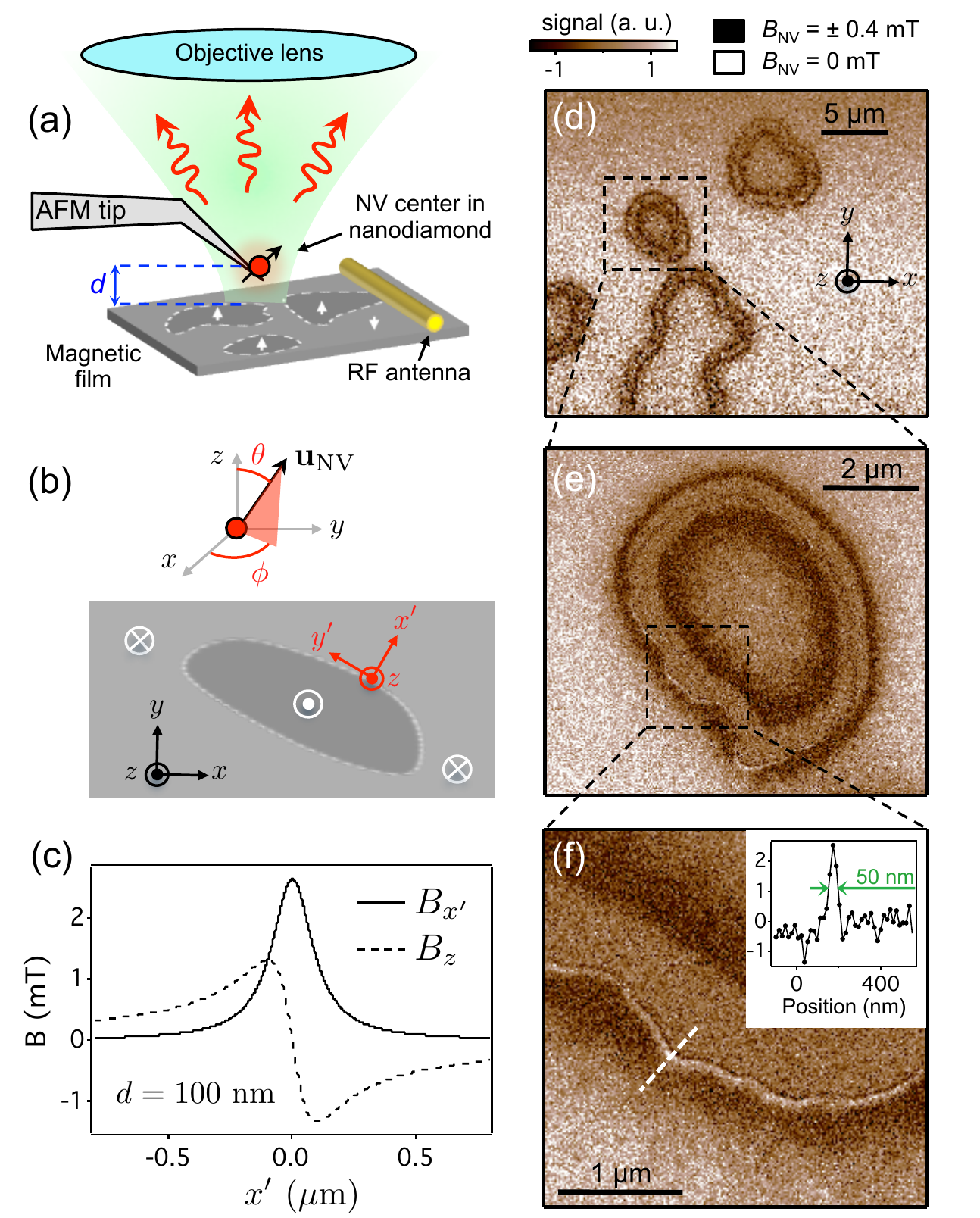}
\caption{(color online) (a)-Schematic of the scanning NV-center microscope. An AFM combined with a confocal microscope and a radiofrequency (RF) antenna enables to measure Zeeman shifts of a single NV defect electron spin placed at the apex of the AFM tip by means of ODMR~\cite{Note}. (b)-Definition of the reference axes. (c)-Stray field components $B_{x^{\prime}}$ and $B_{z}$ calculated at a distance $d=100$ nm as a function the position $x^{\prime}$ across a DW of vanishing width. (d) to (f)-'Dual-iso-B' images measured above a continuous film of Pt/Co(0.6 nm)/AlO$_x$. Positive signal (bright) indicates $B_{\rm NV}=0$ mT and negative signal (dark) indicates $B_{\rm NV}=\pm 0.4$ mT. The inset in (f) shows a linecut across the white dotted line. For these measurements, the NV center projection axis ${\bf u}_{\rm NV}$ is characterized by spherical angles $\theta=138^\circ$ and $\phi=68^\circ$ and the probe-to-sample distance was estimated from independent measurements~\cite{Tetienne2013} to be $d\approx 100$ nm. Integration time per pixel: 100 ms in (d), 150 ms in (e,f).}
\label{Fig1}
\end{center}
\end{figure}
\begin{figure}[b]
\begin{center}
\includegraphics[width=0.42\textwidth]{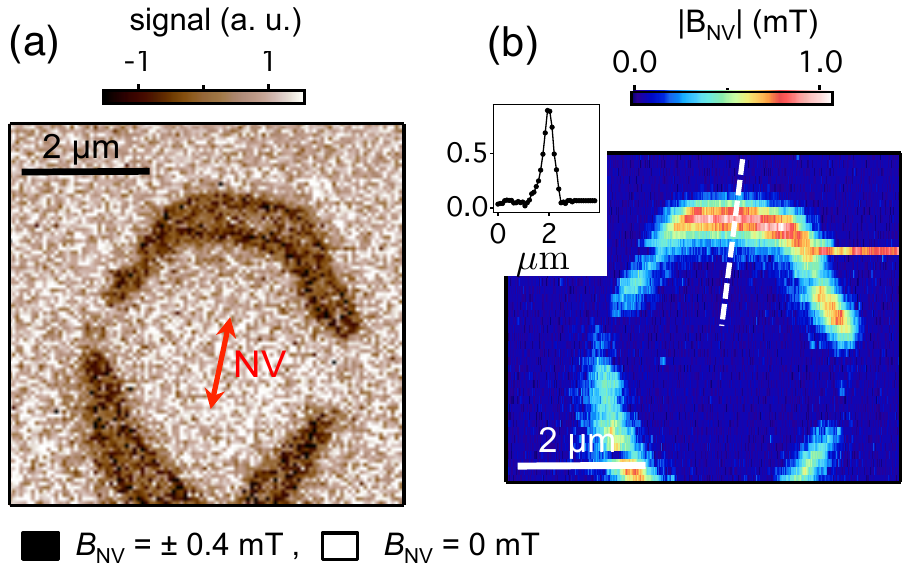}
\caption{(color online) (a)-'Dual iso-B' image of a bubble domain obtained with a NV center's projection axis nearly parallel to the plane ($\theta=88^\circ,\phi=79^\circ$,~see red arrow). (b)-Full stray field distribution $|B_{\rm NV}|$ of the same bubble. Inset: linecut across the DW (white dotted line), revealing a FWHM of $\approx 400$ nm. In these experiments the probe-to-sample distance is $d\approx 200$ nm. Integration time per pixel: 50 ms in (a), 100 ms in (b).}
\label{Fig2}
\end{center}
\end{figure}
\indent As shown in Fig.~\ref{Fig1}(a), the scanning NV-center microscope employs a single NV defect hosted in a diamond nanocrystal, which is attached to the tip of an atomic force microscope (AFM). The NV defect serves as an atomic-size magnetometer by recording Zeeman shift of its electron spin sublevels by means of optically detected magnetic resonance (ODMR)~\cite{Rondin2012}. More precisely, the scanning NV magnetometer provides a quantitative measurement of the component $|B_{\rm NV}|=|{\bf B}\cdot {\bf u}_{\rm NV}|$ of the local magnetic field {\bf B}. Here ${\bf u}_{\rm NV}$ is the NV center's quantization axis which is characterized by the spherical angles ($\theta , \phi$) in the laboratory reference frame ($x,y,z$) [Fig.~\ref{Fig1}(b)]. The magnetic field sensitivity of this magnetometer is $\approx 10$~$\mu$T Hz$^{-1/2}$~[\onlinecite{Rondin2012}]. In this study, we investigate a magnetic sample fabricated from Pt(3 nm)/Co(0.6 nm)/Al(1.6 nm) layers deposited on a thermally oxidized silicon wafer by d.c. magnetron sputtering. After deposition, the aluminium layer was oxidized by exposure to an oxygen plasma. Bubble magnetic domains were nucleated in the continuous film by applying pulses of out-of-plane magnetic field after saturating the sample magnetization in the opposite direction. In the following, we denote the domain wall reference axis ($x^{\prime},y^{\prime},z$), so that $x^{\prime}$ is perpendicular to the DW plane [Fig. \ref{Fig1}(b)].\\
\indent The NV-center magnetic field probe was scanned above the sample and the stray magnetic field was first inferred by using the 'dual-iso-B' imaging mode which provides magnetic images exhibiting two iso-magnetic-field (iso-B) contours~\cite{Note}. For the magnetic images depicted in Fig.~1(d), positive signal (bright) indicates zero field projection along the NV defect quantization axis ($B_{\rm NV}=0$ mT) while negative signal (dark) indicates $B_{\rm NV}=\pm 0.4$ mT, with a projection axis ${\bf u}_{\rm NV}$ measured independently~\cite{Rondin2013}. The stray field is observed to be null everywhere except in ring-like regions: these rings correspond to DWs delimiting magnetic domains with opposite magnetization directions, since a uniformly magnetized film would radiate no stray field.\cite{Thiaville2005} This is in contrast with magneto-optical Kerr microscopy~\cite{Miron2010}, which is sensitive to the magnetization rather than the stray field. Figures~\ref{Fig1}(e)\&(f) show magnified views of a particular bubble domain. Interestingly, a sharp zero-field (bright) line, with a full width at half maximum (FWHM) of $\approx 50$ nm, is clearly visible along the DW in Fig.~\ref{Fig1}(f). It stems from the out-of-plane component of the stray field $B_z$, which vanishes right above the DW.
 \\
\indent To make this point clearer, the in-plane ($B_{x'}$) and out-of-plane ($B_{z}$) components of the stray field above the DW are plotted in Fig.~\ref{Fig1}(c). Neglecting the DW width, and considering a sample thickness $t$ much smaller than the probe-to-sample distance $d$, the stray field of an infinitely long DW writes
\begin{eqnarray} \label{eq:Bx}
{B_{x^{\prime}}} & = & \frac{\mu_0 M_s}{\pi} \left( \frac{td}{x^{\prime 2}+d^2} \right) \ , \  {B_{y^{\prime}}}=0 \\
\label{eq:Bz}
B_{z} & = & -\frac{\mu_0 M_s}{\pi}\left( \frac{tx'}{x'^2+d^2} \right) \ ,
\end{eqnarray}
where $M_s$ is the saturation magnetization. The in-plane component $B_{x'}(x^{\prime})$ therefore follows a Lorentzian profile with a FWHM of $2d$ while the $B_{z}$ component exhibits two extrema at $x^{\prime} = \pm d$ and vanishes at the DW center. For the magnetic images shown in Fig. \ref{Fig1}(d) to (f), $d$ was estimated from independent measurements to be $d\approx 100$~nm~\cite{Tetienne2013} and the NV center projection axis is given by ($\theta=138^\circ$, $\phi=68^\circ$). The measured magnetic field component $B_{\rm NV}$ is therefore a mix of ${B_{x^{\prime}}}$ and $B_{z}$, resulting in complex patterns with a vanishing stray field while crossing the DW [Fig.~\ref{Fig1}(f)]. Analysis of Fig. \ref{Fig1}(e) also shows that the stray field spreads over distances much larger than the DW width (about $6$ nm in such samples~\cite{Miron2010}). Indeed, the wall magnetization contribution is on first approximation negligible so that the apparent width is proportional to the probe-to-sample distance [Eq.~(\ref{eq:Bx}) and (\ref{eq:Bz})]. In addition, the dark iso-B contours in Fig. \ref{Fig1}(d) to (f) (dark areas) correspond to $\pm 0.4$ mT, which is much smaller than the extremum values calculated in Fig. \ref{Fig1}(c). This explains the spacing of about 1 $\mu$m observed between the two main dark lines near a given DW [Fig. \ref{Fig1}(f)]. This effect could be simply overcome by imaging iso-B contours with higher magnetic field magnitudes. However, we stress that since the NV defect probes the field within an atomic size detection volume, it is possible in principle to reconstruct the shape of the DW with a precision down to a few nm, even with $d=100$ nm. This is illustrated by the white contour in Fig.~\ref{Fig1}, which shows fine details over length scales ranging from 100 nm to several $\mu$m. This feature might be of interest to study pinning effects at the nanoscale. \\
\indent For reconstruction purposes, or to make the interpretation of the magnetic images simpler, it would be convenient to measure the stray field component that is either parallel ($B_{x}^{\prime}$) or normal ($B_{z}$) to the sample plane. This can be achieved by selecting a NV defect with the desired orientation~\cite{Rondin2013}. For instance, Fig. \ref{Fig2}(a) shows the 'dual-iso-B' image of a bubble domain obtained with a NV projection axis nearly parallel to the sample plane ($\theta=88^\circ$). As expected, a more symmetrical pattern is observed since the magnetic field probe is only sensitive to the in-plane component of the stray field. For this bubble domain, the full magnetic field distribution was then measured by using a lock-in technique that enables to track the magnetic field value during the scan~\cite{Note,Schoenfeld2011}[Fig. \ref{Fig2}(b)]. As predicted by Eq.~(\ref{eq:Bx}), $|B_{\rm NV}|$ is symmetric with respect to the DW, with a single extremum [see inset in Fig. \ref{Fig2}(b)]. Moreover, the value of the extremum ranges from 0 to $\approx 1$ mT, depending on whether the NV projection axis is parallel ($B_{\rm NV}\approx B_y^{\prime} \approx 0$) or perpendicular ($B_{\rm NV}\approx B_{x'}$) to the DW.
\begin{figure}[t]
\begin{center}
\includegraphics[width=0.45\textwidth]{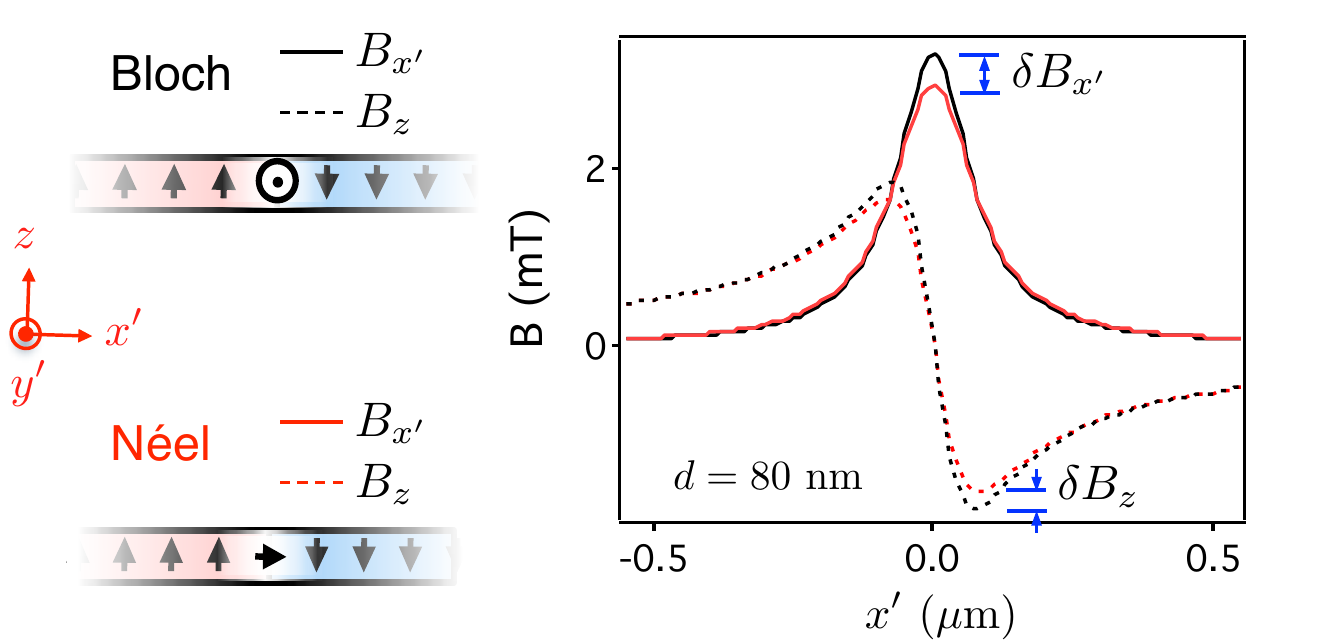}
\caption{(color online) In-plane ($B_{x'}$) and out-of-plane ($B_{z}$) stray field components calculated at a distance $d=80$ nm as a function of $x'$ across the DW whose center is at $x'=0$. The DW is initialized either in a Bloch or in a N{\'e}el-type configuration. The parameters used in OOMMF software are $M_s=1.1$ MA/m, $t=0.6$~nm, exchange constant $A = 16$ pJ/m and anisotropy constant $K = 1.27$ MJ/m$^3$.}
\label{Fig3}
\end{center}
\end{figure}

These measurements demonstrate that NV-center microscopy is a valuable tool for the study of ultrathin ferromagnets. A step beyond is to give information on the wall micromagnetic structure. We anticipate that quantitative stray field measurements similar to those reported in this letter should enable to distinguish between a N\'eel or Bloch DW in Pt/Co/AlO$_x$ and other ultrathin ferromagnets. Indeed, while Eqs. (\ref{eq:Bx}) and (\ref{eq:Bz}) assumed an abrupt DW, the numerical calculations shown in Fig.~\ref{Fig3} take into account the fine DW micromagnetic structure, which can be either of Bloch or N{\'e}el type. This calculations were performed using OOMMF software~\cite{oommf} with a magnetization cell size $2.5\times 2.5\times 0.6 \  {\rm nm}^{3}$. Interestingly, the stray field above the DW slightly differs for these distinct structures, owing to the contribution of the in-plane magnetization at the center of a N{\'e}el DW [Fig.~\ref{Fig3}]. For instance, at a distance $d=80$~nm, which can be achieved routinely with our instrument~\cite{Rondin2013}, $B_{x'}$ would reach $2.9$~mT above a N{\'e}el DW (with the chirality depicted in Fig.~\ref{Fig3}) against $3.3$ mT for a Bloch DW. Discriminating these values is by far within the operation range of scanning NV-center microscopy. Quantitative stray field imaging with a single NV defect could therefore solve a controversial issue in the field of nanomagnetism, the stabilization of N\'eel walls by the Dzyaloshinskii-Moriya interaction in ultrathin magnetic films being still under debate~\cite{Thiaville2012}.\\
\indent  This work was supported by the Agence Nationale de la Recherche (ANR) through the project Diamag, and by C'Nano \^Ile-de-France (contracts Magda and Nanomag).


\begin{thebibliography}{50}

\bibitem{Miron2010}
I. M. Miron {\it et al.}, Nat. Mater. {\bf 9}, 230 (2010).

\bibitem{Miron2011}
I. M. Miron {\it et al.}, Nature {\bf 476}, 189 (2011).

\bibitem{LiuScience2012}
L. Liu, C.-F. Pai, Y. Li, H. W. Tseng, D. C. Ralph, and R. A. Buhrman, Science {\bf 336}, 555 (2012).

\bibitem{Garello2013}
K. Garello {\it et al.}, Nat. Nano. {\bf 8}, 587 (2013).

\bibitem{Thiaville2012}
A. Thiaville, S. Rohart, E. Ju\'e, V. Cros, A. Fert, Europhys. Lett. {\bf 100}, 57002 (2012).

\bibitem{Ryu2013}
K.-S. Ryu, L. Thomas, S.-H. Yang, and S. Parkin, Nat. Nano. {\bf 8}, 527 (2013).

\bibitem{Emori2013}
S. Emori, U. Bauer, S.-M. Ahn, E. Martinez, and G. S. D. Beach, Nat. Mater. {\bf 12}, 611 (2013).



\bibitem{Heyne2008}
L. Heyne {\it et al.}, J. Appl. Phys. {\bf 103}, 07D928 (2008).

\bibitem{Im2009}
M.-Y. Im, L. Bocklage, P. Fischer, and G. Meier, Phys. Rev. Lett. {\bf 102}, 147204 (2009).

\bibitem{Vogel2012}
J. Vogel {\it et al.}, Phys. Rev. Lett  {\bf 108}, 247202 (2012).

\bibitem{Klaui2005}
M. Klaui {\it et al.}, Phys. Rev. Lett. {\bf 95}, 026601 (2005).

\bibitem{Biziere2013}
N. Biziere, C. Gatel, R. Lassalle-Balier, M. C. Clochard, J. E. Wegrowe, E. Snoeck, Nano Lett. {\bf 13}, 2053 (2013).

\bibitem{Yamaguchi2004}
A. Yamaguchi, T. Ono, S. Nasu, K. Miyake, K. Mibu, T. Shinjo, Phys. Rev. Lett. {\bf 92}, 077205 (2004).


\bibitem{Meckler2009}
S. Meckler {\it et al.}, Phys. Rev. Lett  {\bf 103}, 157201 (2009).

\bibitem{Chen2013}
G. Chen, {\it et al.} Phys. Rev. Lett. {\bf 110}, 177204 (2013).

\bibitem{Taylor2008}
J. M. Taylor {\it et al.}, Nat. Phys. {\bf 4}, 810 (2008).

\bibitem{Balasubramanian2008}
G. Balasubramanian {\it et al.}, Nature {\bf 455}, 648 (2008).

\bibitem{Rondin2012}
L. Rondin {\it et al.}, Appl. Phys. Lett. {\bf 100}, 153118 (2012).

\bibitem{Maletinsky2012}
P. Maletinsky {\it et al.}, Nat. Nano. {\bf 7}, 320 (2012).


\bibitem{Maze2008}
J. R. Maze {\it et al.}, Nature {\bf 455}, 644 (2008).

\bibitem{Grinolds2013}
M. S. Grinolds {\it et al.}, Nature Phys. {\bf 9}, 215 (2013).

\bibitem{Rondin2013}
L. Rondin {\it et al.}, Nat. Commun. {\bf 4}, 2279 (2013).

\bibitem{Tetienne2013}
J.-P. Tetienne {\it et al.}, preprint arXiv:1309.2171 (2013).

\bibitem{LeSage2013}
D. Le Sage {\it et al.}, Nature {\bf 496}, 486 (2013).

\bibitem{Note}
Details about the experimental setup and the magnetic field imaging methods can be found in Ref.~\cite{Rondin2012,Rondin2013}.

\bibitem{Thiaville2005}
A. Thiaville, J. Miltat, J.M. Garcia, in Magnetic Force Microscopy, H. Hopster and H.P. Oepen eds. (Springer-Verlag), p. 225 (2005)

\bibitem{Schoenfeld2011}
R. Schoenfeld, W. Harneit, Phys. Rev. Lett. {\bf 106}, 030802 (2011).

\bibitem{oommf}
M. J. Donahue and D. G. Porter, OOMMF User's Guide, \textit{http://math.nist.gov/oommf} (1999).

\end{thebibliography}
\end{document}